\documentclass[aps,pre,twocolumn,groupedaddress,showpacs]{revtex4-1}
\usepackage[T1]{fontenc}
\usepackage[utf8]{inputenc}
\usepackage{graphicx}
\usepackage{dcolumn}
\usepackage{bm}
\usepackage{amsmath}
\usepackage{amssymb}
\usepackage{dsfont}
\newcommand{\indykator}[1]{\mathds{1}_{\left\{#1\right\}}}

\usepackage{lipsum} 
\usepackage{float}

\begin{document}

\preprint{APS/123-QED}

\title{In search for the social hysteresis -- the symmetrical threshold model with
independence on Watts–Strogatz graphs}

\author{Bart\l{}omiej Nowak}
 \altaffiliation[]{bartlomiej.nowak@pwr.edu.pl}
\author{Katarzyna Sznajd-Weron}%
 \email{katarzyna.weron@pwr.edu.pl}
\affiliation{%
 Department of Theoretical Physics, Wroclaw University of Science and Technology, Wrocław, Poland.
}%

\date{\today}

\begin{abstract}
We study the homogeneous symmetrical threshold model with independence (noise) by pair approximation and Monte Carlo simulations on Watts–Strogatz graphs. The model is a modified version of the famous Granovetter's threshold model: with probability $p$ a voter acts independently, i.e. takes randomly one of two states $\pm 1$; with complementary probability $1-p$, a voter takes a given state, if sufficiently large fraction (above a given threshold $r$) of individuals in its neighborhood is in this state. We show that the character of the phase transition, induced by the noise parameter $p$, depends on the threshold $r$, as well as graph's parameters. For $r=0.5$ only continuous phase transitions are observed, whereas for $r>0.5$ also discontinuous phase transitions are possible. The hysteresis increases with the average degree $\langle{k}\rangle$ and the rewriting parameter $\beta$. On the other hand, the dependence between the width of the hysteresis and the threshold $r$ is non-monotonic. The value of $r$, for which the maximum hysteresis is observed, overlaps pretty well the size of the majority used for the descriptive norms in order to manipulate people within social experiments. We put results obtained within this paper in a broader picture and discuss them in the context of two other models of binary opinions, namely the majority-vote and the $q$-voter model.
\end{abstract}

\maketitle


\section{\label{sec:Introduction}Introduction}
It is not surprising that binary opinion models are extremely popular among sociophysicists, given that the $1/2$-spin Ising model is not only one of the most popular models of theoretical physics, but also absolutely fundamental for the theory of phase transitions. However, what is probably more surprising, the binary-choice models have received considerably more theoretical attention than other choice models among social psychologists, sociologists and economists \cite{Wat:Dod:17,Jed:Szn:19}. One of the most important class of such models are the threshold models \cite{Wat:02,Gra:Li:19} taking roots in the pioneering paper by Granovetter \cite{Gra:78}. 

The idea behind these models is extremely simple -- an agent takes state $1$ (which can be interpreted as agree, adopt the innovation, join the riot, etc.) if sufficiently large fraction (above a given threshold) of people in his neighborhood is in state $1$. Originally model has been investigated under the assumption of perfect mixing (all-to-all interactions). However, in 2002 Watts has adapted Granovetter’s threshold model to a network framework \cite{Wat:02}. We will use the same approach here and therefore individuals will be influenced only by the nearest neighbors, i.e. interactions will take place only between agents that are directly linked.

There are two important differences between the Watts threshold model and other models of binary opinions, such as the Galam model \cite{Gal:90,Gal:08,Gal:12}, the majority-vote (MV) \cite{Lig:85,Tom:Oli:San:91,Oli:92,Lim:Mal:06,San:Lim:Mal:10,Vie:Cro:16,Che:etal:17,Fro:Fro:17,Kra:18,Kra:Gra:19,Enc:etal:18,Enc:etal:19}, the $q$-voter (qV) \cite{Cas:Mun:Pat:09,Nyc:Szn:Cis:12,Mor:etal:13,Mob:15,Jav:Squ:15,Mel:Mob:Zia:16,Mel:Mob:Zia:17,Jed:17} or the threshold $q$-voter (TqV) model \cite{Nyc:Szn:13,Vie:Ant:18,Nyc:etal:18,Vie:etal:20}. The first difference, often considered as the most important, is the heterogeneity -- each agent is described by an individual threshold and therefore some agents adopt a new state very easily, whereas others don't \cite{Wat:02}. The second difference, that should be particularly important for physicists, is the lack of the up-down symmetry. Once an agent adopts a state $1$ it cannot go back to the previous one. To make the threshold model comparable with other binary opinion models, we have introduced recently the homogeneous symmetrical threshold model \cite{Now:Szn:19}.  Here we will call this model simply symmetrical threshold (ST) model for brevity. 

Previously, we have studied two versions of the ST model, each with a different type of nonconformity (anticonformity or independence) on the complete graph \cite{Now:Szn:19}. Therefore we were able to obtain exact analytical results within the mean-field approach. Analogously as in other models of binary opinions, the introduction of nonconformity, whether in the form of anticonformity or in the form of independence, resulted in the appearance of the agreement--disagreement phase transitions. We have shown, that for the threshold $r=0.5$, which corresponds to the majority-vote model, the phase transition is continuous, whereas for $r>0.5$ a discontinuous phase transitions appear within the model with independence. For the model with anticonformity phase transitions are continuous for an arbitrary value of $r$. Similar phenomenon has been observed previously for the $q$-voter model -- within the model with anticonformity only continuous phase transitions are observed, whereas within the model with independence (known also as the nonlinear noisy voter model) a discontinuous phase transitions appear for $q > 5$ \cite{Nyc:Cis:Szn:12,Jed:17,Per:etal:18}.

In this paper we focus on the ST model with independence, because it occurs that the hysteresis and tipping points, two signatures of a discontinuous phase transitions, are common features of complex social systems \cite{Sch:Wes:Bro:03,Val:Now:Rea:17,Cen:etal:18}. We study the model on Watts-Strogatz (WS) graph \cite{Watt:Str:98} because it allows to tune the structure from (1) the complete graph, for which the mean-field approximation gives exact result, through (2) random graphs for which the pair approximation should work properly, to (3) small-world networks which resembles the basic features of the real social networks. Because it has been shown recently that the size of the hysteresis may depend on the graph's properties, we focus on this issue and check to what extend results found within the MV model and the qV model are universal \cite{Vie:Cro:16,Che:etal:17,Enc:etal:18,Enc:etal:19,Jed:17,Abr:Szn:20}. 

\section{\label{sec:Model}Model}
We consider a system of $N$ individuals placed in the nodes of an arbitrary graph. Each node represents exactly one individual (interchangeably called \textit{an agent}, \textit{a spin}, or \textit{a voter}). We consider a model of binary opinions/believes/decisions and thus each voter at time $t$ is described by a binary dynamical variable $S_i(t) = \pm 1 (\uparrow/\downarrow)$. At each elementary update $\Delta t$:
\begin{enumerate}
	\item a site $i$ is randomly chosen from the entire graph,
	\item an agent at site $i$ acts independently with probability $p$, i.e. changes its opinion to the opposite one $S_{i}(t + \Delta t) = -S_{i} (t)$ with probability $\frac{1}{2}$,
	\item with complementary probability $1-p$ it conforms to its $k_i$ neighbors if the fraction of its neighbors in the same state is larger than $r$:
	\begin{enumerate}
		\item $S_{i}(t + \Delta t) = 1$ if more than $rk_i$ neighbors are in the state $1$ or
		\item $S_{i}(t + \Delta t) = - 1$ if more than $rk_i$ neighbors are in the state $-1$. 
	\end{enumerate}
\end{enumerate}
As usual, a single Monte Carlo step consists of $N$ updates, i.e. $\Delta t = 1/N$, which means that
one time unit corresponds to the mean update time of a single individual. 
Under the above algorithm the following changes are possible in the system:
\begin{equation}
\begin{split}
\underbrace{\uparrow \uparrow \dots \uparrow}_{>\lfloor{rk_{i}}\rfloor} \Downarrow &\stackrel{1-p}{\longrightarrow} \underbrace{\uparrow \uparrow \dots \uparrow}_{>\lfloor{rk_{i}}\rfloor} \Uparrow, \\
\underbrace{\downarrow \downarrow \dots \downarrow}_{>\lfloor{rk_{i}}\rfloor} \Uparrow &\stackrel{1 - p}{\longrightarrow} \underbrace{\downarrow \downarrow \dots \downarrow}_{>\lfloor{rk_{i}}\rfloor} \Downarrow ,\\
\underbrace{\dots\dots\dots}_{\substack{\text{any} \\ \text{configuration}}} \Uparrow &\stackrel{p/2}{\longrightarrow} \underbrace{\dots\dots\dots}_{\substack{\text{any} \\ \text{configuration}}} \Downarrow, \\
\underbrace{\dots\dots\dots}_{\substack{\text{any} \\ \text{configuration}}} \Downarrow &\stackrel{p/2}{\longrightarrow} \underbrace{\dots\dots\dots}_{\substack{\text{any} \\ \text{configuration}}} \Uparrow, \\
\end{split}
\label{eq:equation1}
\end{equation}
where $\Downarrow$ and $\Uparrow$ denotes states of a target agent, and $\lfloor{rk_i}\rfloor$ is the floor function of $rk_i$. In any other situation, the state of the system does not change.

In the Watts threshold model flipping from $\uparrow$ to $\downarrow$, was forbidden \cite{Wat:02}. Therefore, the model was asymmetrical on contrary to the majority--vote or the $q$-voter. 

In the original threshold model an arbitrary value of $r \in [0,1]$ is possible, which is a reasonable assumption for the asymmetrical model describing the adoption to the new state. In the symmetrical case, the situation for $r<0.5$ is less obvious. It can be easily seen within the following example: let the threshold $r<0.5$ and the neighborhood of a target voter consists of $50\%$ positive and $50\%$ negative agents. It means that both opinions (positive and negative) could be adopted by the voter. Which one should be chosen in such a situation? 

There are several possibilities to solve the above ambiguity, e.g. we can assume that: (1) a voter prefers to change opinion and therefore will always change it to the opposite one whenever possible \cite{Vie:Ant:18,Vie:etal:20}, (2) a voter prefers to keep an old opinion; this assumption overlaps $r \ge 0.5$ \cite{Nyc:Szn:13,Now:Szn:19} (3) a voter makes a random decision to flip or keep an old state. Each of these scenarios can be used. However, for modeling opinion/belief formation the second one, i.e. $r \ge 0.5$, seems to be the most justified from the social point of view \cite{Nyc:etal:18}.

\section{\label{sec:PA}Analytical approach within pair approximation}
Our analytical approach is based on the pair approximation (PA), an improved version of the standard mean-field approximation (MFA), which has been already applied to various binary--state dynamics on complex networks \cite{Gle:13,Jed:17,Per:etal:18a}. 

Because at each elementary update only one voter can change his opinion thus the number of agents with positive opinion $N_{\uparrow}$ increases or decreases by $1$ or remains constant. As in \cite{Nyc:Cis:Szn:12} we denote by $c=N_{\uparrow}/N$ the concentration of the positive opinion, which in an elementary time step
increases or decreases by $\frac{1}{N}$ or remains constant. We also denote transition probabilities as in \cite{Nyc:Szn:Cis:12}:
\begin{equation}
\begin{aligned}
\gamma^{+} &= Prob\left(c(t+\Delta t) = c(t) + \frac{1}{N}\right), \\
\gamma^{-} &= Prob\left(c(t+\Delta t) = c(t) - \frac{1}{N}\right), \\
\gamma^{0} &= Prob\left(c(t+\Delta t) = c(t) \right) = 1- \gamma^{+} - \gamma^{-}.
\end{aligned}
\label{eq:equation2}
\end{equation}
For $N \rightarrow \infty$ we can safely assume that random variable $c$ localize to the expectation value and we get the following continuous time dynamical system:
\begin{equation}
    \frac{d c}{d t} = \gamma^{+} - \gamma^{-},
\label{eq:equation3}
\end{equation}
in the rescaled time units $t$. The simplest and the most popular approach under which formulas for transition probabilities $\gamma^{\pm}$ can be derived analytically is the simple mean-field approach \cite{Cas:Mun:Pat:09,Nyc:Szn:Cis:12,Mor:etal:13,Nyc:Szn:13,Nyc:etal:18,Vie:Ant:18,Now:Szn:19}. It gives very good agreement for the complete graph, but rarely for more complicated structures, because it neglects all fluctuations in the system by assuming that the local concentration of spins up is equal to the global one. 

Another method, which works particularly well for random graphs with low clustering coefficient, is the pair approximation. Within PA we describe the system by two differential equations -- one for the time evolution of the concentration $c$ of spins up and the second one for the time evolution of the concentration $b$ of active bonds/links (bonds between two opposite spins) \cite{Gle:13,Jed:17,Jed:Szn:19}:
\begin{align}
\frac{d c}{d t}=&-\sum_{j\in\{1,-1\}}c_j\sum_k P(k)\sum_{i=0}^{k}{k\choose i}\theta_j^i(1-\theta_j)^{k-i} \nonumber \\
&\times f(i,r,k)j, \label{eq:equation4}\\
\frac{d b}{d t}=&\frac{2}{\langle k\rangle}\sum_{j\in\{1,-1\}}c_j\sum_k P(k)\sum_{i=0}^{k}{k\choose i}\theta_j^i(1-\theta_j)^{k-i} \nonumber \\
&\times f(i,r,k)(k-2i) \label{eq:equation5}, 
\end{align}
where: 
\begin{itemize}
	\item $c_j$ is the concentration of spins in state $j =\pm 1$ and thus $c_{1}=c$, $c_{-1}=1-c$,
	\item $P(k)$ is the degree distribution of a graph and $\langle k \rangle$ is the average node degree,
	\item $\theta_j$ is the conditional probability of selecting a node that is in the opposite state to its neighbor in a state $j$, which is equivalent to the probability of choosing an active link from all links of a node in state $j$ and can be approximated by \cite{Jed:17,Jed:Szn:19}:
	\begin{equation}
	\theta_j = \frac{b}{(2c_j)},
	\label{eq:theta}
	\end{equation} 
	\item $f(i,r,k)$ is the flipping probability, i.e. the probability that a node in state $j$ changes its state under the condition that exactly $i$ from its $k$ links are active.
\end{itemize}

Within our version of the threshold model, a voter flips with probability $1/2$ due to the independence, which takes place with probability $p$ or due to the conformity, which takes place with probability $1-p$ if more than $\lfloor{rk}\rfloor$ of its nearest neighbors are in the opposite state and thus:
\begin{equation}
    f(i,r,k) = \frac{p}{2} + (1-p)\indykator{i > \lfloor{rk}\rfloor},
\label{eq:equation6}
\end{equation}
where $\indykator{i > \lfloor{rk}\rfloor}$ is the indicator function, i.e. gives $1$ for $i > \lfloor{rk}\rfloor$ and $0$ otherwise.

We consider the model on the WS graph and thus the degree probability $P(k)$ equals \cite{Bar:Pas:10}:
\begin{align}
P(k)&=\sum_{n=0}^{f(k, K)}\left(\begin{array}{c}
{K / 2} \\ {n}
\end{array}\right)(1-\beta)^{n} \beta^{K / 2-n} \nonumber \\
&\times \frac{(\beta K / 2)^{k-K / 2-n}}{(k-K / 2-n) !} e^{-\beta K / 2}. \label{eq:equation7}
\end{align}
PA works properly for small clustering coefficients which correspond to large values of $\beta$. Moreover, under the assumption $\beta \rightarrow 1$, calculations simplify substantially, since Eq. (\ref{eq:equation7}) reduces to:
\begin{equation}
    P(k) = \frac{(K/2)^{k-K/2}}{(k - K/2)!}e^{-K/2}.
\label{eq:equation8}
\end{equation}
Therefore, we take in further calculations $P(k)$ given by Eq. (\ref{eq:equation8}).

After inserting $f(i,r,k)$, given by Eq. (\ref{eq:equation6}), into Eqs. (\ref{eq:equation4}) -- (\ref{eq:equation5}) we obtain:
\begin{align}
\frac{d c}{d t}=&-\sum_{j\in\{1,-1\}} c_j \sum_k P(k)\Bigg[\frac{jp}{2} + \nonumber \\  
&+ j(1-p)\sum_{i=\lfloor{rk}\rfloor+1}^{k}{k\choose i}\theta_j^i(1-\theta_j)^{k-i}\Bigg], \label{eq:equation9}\\
\frac{d b}{d t}=&\frac{2}{\langle k\rangle}\sum_{j\in\{1,-1\}}c_j \sum_k P(k)\Bigg[pk \left(\frac{1}{2}-\theta_j \right) + \nonumber\\
&+ (1-p)\sum_{i=\lfloor{rk}\rfloor+1}^{k}{k\choose i}\theta_j^i(1-\theta_j)^{k-i}(k-2i)\Bigg]. \label{eq:equation10}
\end{align}
The steady states can be obtained by solving equations:
\begin{eqnarray}
\frac{d c}{d t} &=& 0, \label{eq:stat:c}\\
\frac{d b}{d t} &=& 0.
\label{eq:stat:b}
\end{eqnarray}
Analogously as for the $q$-voter model with independence, we are not able to solve above equations explicitly  but we can obtain inverse relation $p=p(c)$, instead of $c=c(p)$ \cite{Nyc:Cis:Szn:12}. For the concentration of active bonds we can present only implicit solution. 

One solution of Eq. (\ref{eq:stat:c}), namely $c = 1/2$, is straightforward because it is seen that for this value the right side of Eq. (\ref{eq:equation9}) equals to zero, i.e. point $c=1/2$ is the fixed point for all values of $p$. On the other hand, the right side of Eq. (\ref{eq:equation10}) is nonzero at $c = 1/2$, thus from Eq. (\ref{eq:stat:b}) for $c = 1/2$ we can derive the relation $p(b)$: 
\begin{equation}
p = \frac{
	\sum\limits_{k}P(k)\sum\limits_{i=\lfloor{rk}\rfloor+1}^{k}{k\choose i}b^i(1-b)^{k-i}(k-2i)}{- \langle{k}\rangle\left(\frac{1}{2} - b\right) +
	\sum\limits_{k}P(k)\sum\limits_{i=\lfloor{rk}\rfloor+1}^{k}{k\choose i}b^i(1-b)^{k-i}(k-2i)
} 
\label{eq:equation13}
\end{equation}
We see that $b \rightarrow 0$ gives $p = 0$ and $b \rightarrow 1/2$ gives $p = 1$.

\begin{figure}
	\includegraphics[width=.48\textwidth]{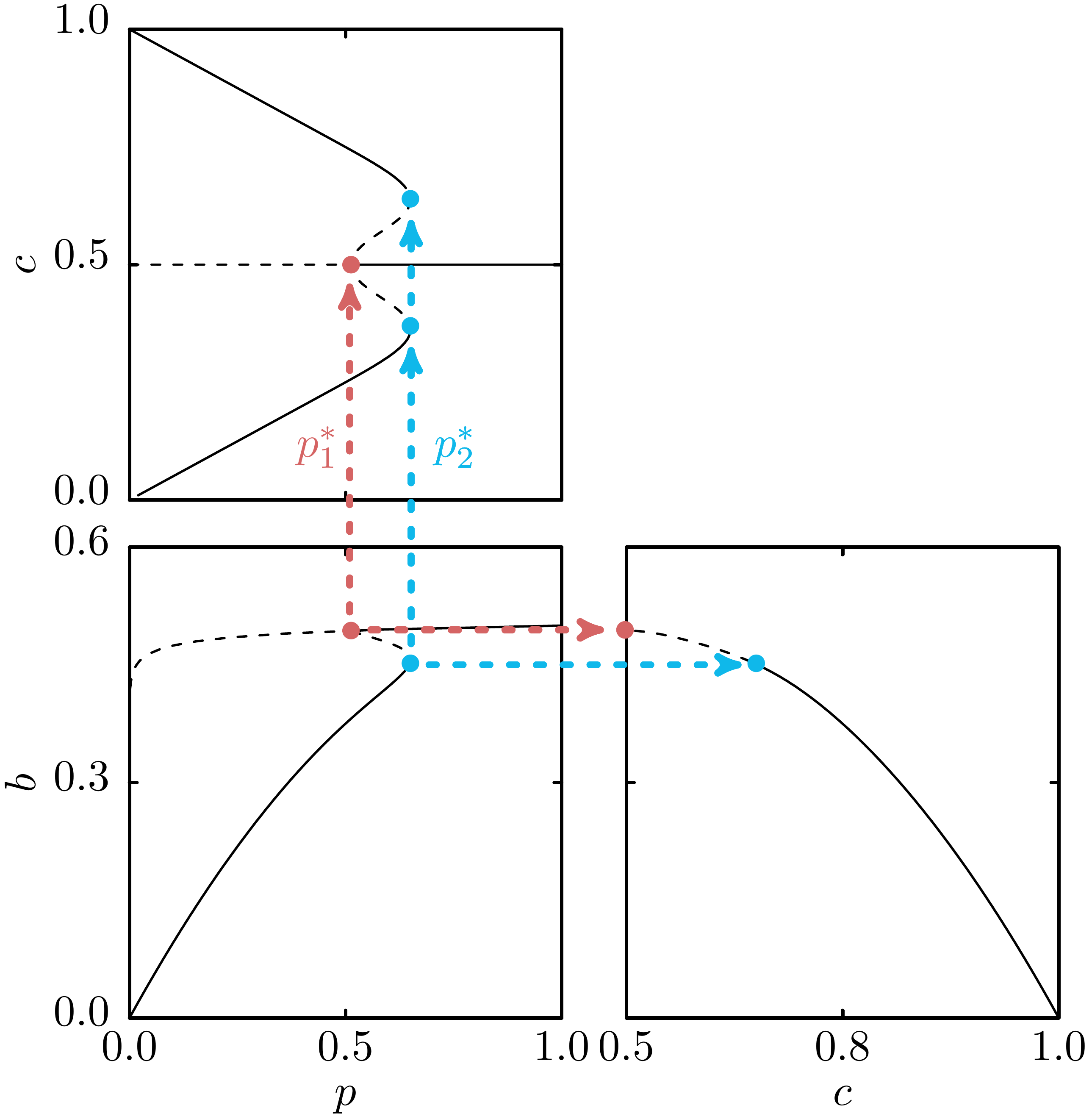}
	\caption{Dependencies between the stationary value of the concentration of spins up $c$, and active bonds $b$ and the noise $p$ obtained within PA for sample values of parameters $\langle{k}\rangle=80$ and $r=0.6$. Results are presented in three phase-space projections: $(c,p),(b,p)$ and $(b,c)$. For $p<p_1^*$ the only stable solution is the ordered phase, in which the symmetry between $\uparrow$ and $\downarrow$ states is broken, whereas for $p>p_2^*$ the only stable solution is the disordered phase.}
	\label{fig:bif}
\end{figure}

\noindent To show the behavior of the system for $c \neq 1/2$ we insert Eq. (\ref{eq:equation9}) to  Eq. (\ref{eq:stat:c}), which allows to derive the relation:
\begin{widetext} 
\begin{equation}
p = 
\frac{\sum\limits_{k}P(k)\sum\limits_{i=\lfloor{rk}\rfloor+1}^{k} {k\choose i} \left(c\theta_{\uparrow}^i(1-\theta_{\uparrow})^{k-i} - (1-c)\theta_{\downarrow}^i(1-\theta_{\downarrow})^{k-i}\right)}
{\frac{1}{2} - c + \sum\limits_{k}P(k)\sum\limits_{i=\lfloor{rk}\rfloor+1}^{k}{k\choose i}\left(c \theta_{\uparrow}^i(1-\theta_{\uparrow})^{k-i} - (1-c)\theta_{\downarrow}^i(1-\theta_{\downarrow})^{k-i} \right)},
\label{eq:equation11}
\end{equation}
\end{widetext} 
where we denoted $\theta_{1/-1}$ by $\theta_{\uparrow/\downarrow}$ for clarity. Note that the above equation is in fact the relation $p = p(c,b)$, because both $b$ and $c$ are implicitly included in $\theta_{\uparrow}$ and $\theta_{\downarrow}$ according to Eq. (\ref{eq:theta}). Thus, to solve the above equation we need the relation $b=b(c)$, which can be obtained by inserting the above equation into Eq. (\ref{eq:stat:b}):
\begin{widetext}
\begin{equation}
    \begin{aligned}
        0 &= \sum\limits_{k}P(k)\sum\limits_{i=\lfloor{rk}\rfloor+1}^{k}{k\choose i} \Bigg[
        c\theta_{\uparrow}^i(1-\theta_{\uparrow})^{k-i} \Big( \langle{k}\rangle (1-2b) + (1-2c)(k-2i) \Big) + \\ 
        &+(1-c)\theta_{\downarrow}^i(1-\theta_{\downarrow})^{k-i} \Big( (1-2c)(k-2i) - \langle{k}\rangle (1-2b)\Big) \Bigg].
    \end{aligned}
\label{eq:equation12}
\end{equation}
\end{widetext} 
As we have noticed above, Eq. (\ref{eq:equation11}) gives the relation $p = p(c,b)$, which can be plotted in 3 different planes, as shown in Fig. \ref{fig:bif}. There are two critical points, seen in this plot: (1) $p = p^*_1$, in which solution $c=1/2$ losses stability (so called lower spinodal), (2) $p = p^*_2$, in which solution $c=c(p) \ne 1/2$, given by Eq. (\ref{eq:equation11}), loses stability. There are several possibilities to calculate $p = p^*_1$ \cite{Nyc:Szn:Cis:12,Jed:17,Nyc:etal:18}. Here we use method based on the observation that  $p=p^*_1$ corresponds to the point $c=1/2$ in the relation $b=b(c)$ (right bottom panel Fig. \ref{fig:bif}). Therefore, first we take a limit $c \rightarrow 1/2$ in Eq. (\ref{eq:equation12}), which gives:
\begin{widetext}
\begin{equation}
\begin{aligned}
0 &= \sum\limits_{k} P(k) \sum\limits_{i=\lfloor{rk}\rfloor+1}^{k}{k\choose i}b^i(1-b)^{k-i} \Bigg[
k - \langle{k}\rangle(1-2b)\left(1+\frac{kb}{1-b}\right)+ \\
&-2i + \langle{k}\rangle(1-2b)\left(1+\frac{b}{1-b}\right)i \Bigg].
\end{aligned}
\label{eq:equation14}
\end{equation}
\end{widetext}
and then derive $b$ from the above equation. Finally we insert this value of $b$ to Eq.(\ref{eq:equation13}), which gives $p=p^*_1$. The upper spinodal, i.e. point $p = p^*_2$, where $p=p(c)$ has two maxima (see Fig. \ref{fig:bif}), can be calculated numerically from Eq.(\ref{eq:equation11}) by taking a maximum value of $p$.

\begin{figure}
	\includegraphics[width=0.48\textwidth]{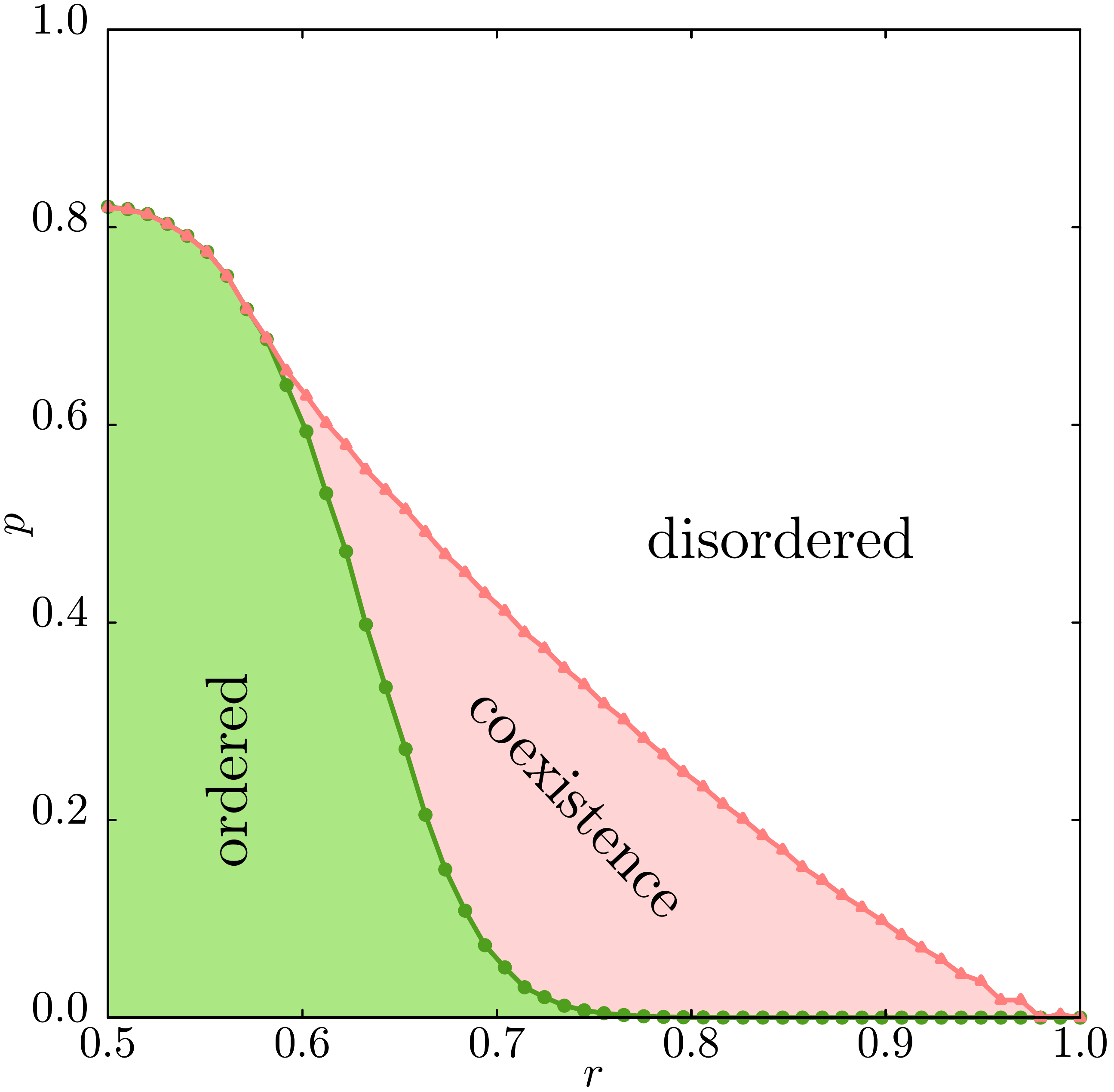}
	\caption{Phase diagrams for the average degree $\langle{k}\rangle = 50$ and $r=0.6$. Lines with represents spinodals obtained within PA from Eqs. (\ref{eq:equation11}) - (\ref{eq:equation14}) i.e., limits of the region with metastability, in which the final state depends on the initial one.}
	\label{fig:phase_diagram}
\end{figure}

\section{\label{sec:discussion} Discussion of the pair approximation results}
It was shown that for the majority-vote model with inertia there are two ingredients responsible for the discontinuous phase transitions: (1) the level of inertia and (2) the average node degree $\langle{k}\rangle$ \cite{Che:etal:17,Enc:etal:18}. Similarly, for the $q$-voter model (1) the size of the influence group $q$ and (2) $\langle{k}\rangle$ and are key factors influencing the type of the phase transition \cite{Nyc:Cis:Szn:12,Jed:17,Abr:Szn:20}. The question is if the same can be seen within the ST model.

The first ingredient influencing the phase transition was studies already in the previous paper within the mean-field approach \cite{Now:Szn:19}. We have observed continuous phase transitions for $r=0.5$ and discontinuous for $r>0.5$. We have obtained similar result within PA, as shown in Fig.\ref{fig:phase_diagram}: for small values of $r$ we observe a continuous, whereas for larger $r$ a discontinuous phase transition. This result is similar to results obtained within the MV model with inertia and the qV model. In both models discontinuous phase transitions were observed only for the sufficiently large value of inertia $\theta$ \cite{Che:etal:17,Enc:etal:18} or the large size of the influence group $q$ \cite{Nyc:Cis:Szn:12,Jed:17}. It should be noticed that both the large size of the influence group $q$ and the high threshold $r$ corresponds to the high value of inertia:
\begin{description}
	\item[qV model] it is unlikely to find a unanimous group of size $q$ if $q$ is large,
	\item[ST model] it is unlikely to find a fraction of agents in the same state larger than $r$ if $r$ is large.
\end{description}
Therefore, in both cases a voter is unlikely influenced by neighbors, i.e. its inertia is larger.

\begin{figure}[t]
	\includegraphics[width=.48\textwidth]{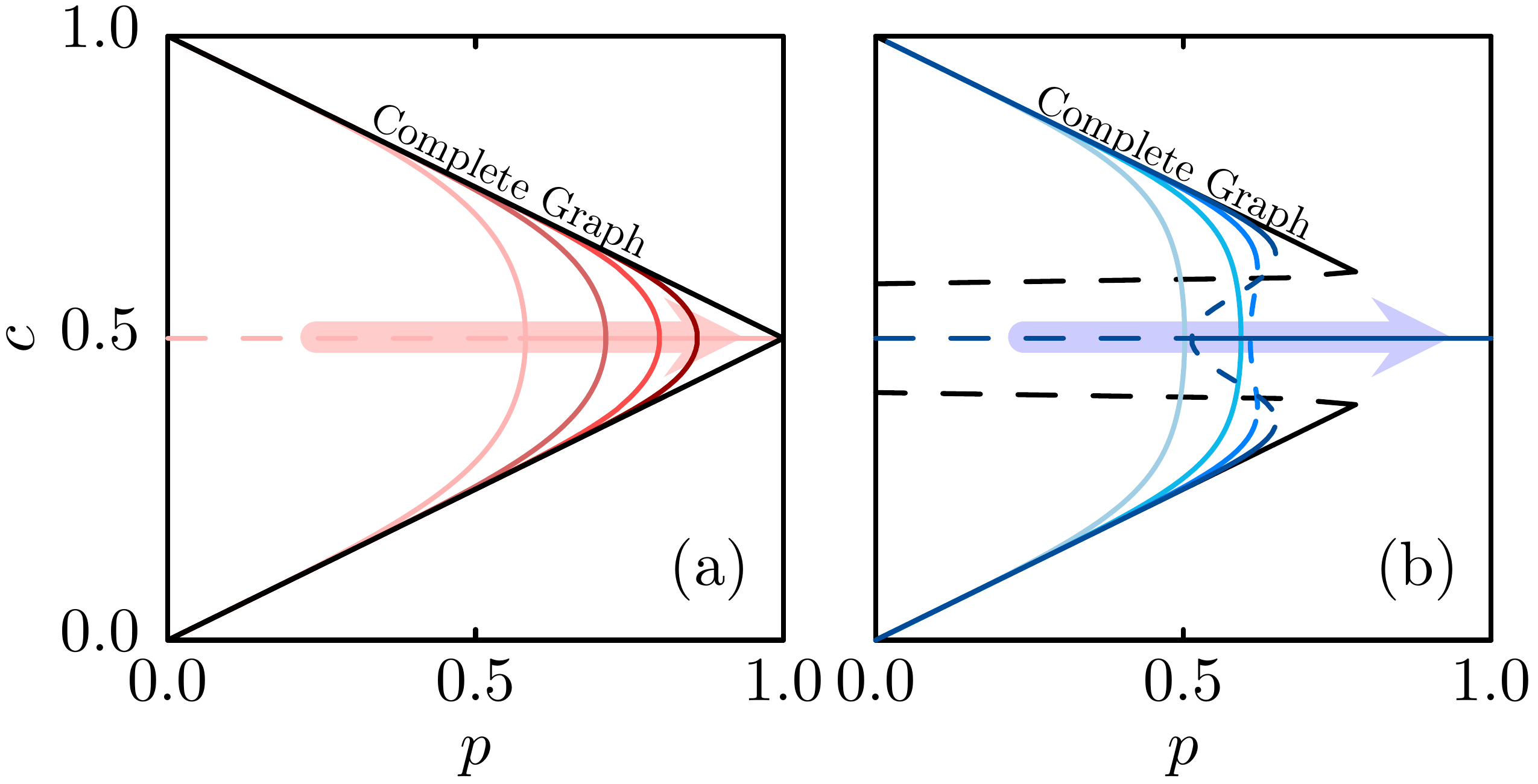}
	\caption{Dependence between the stationary concentration of spins up $c$ and the noise $p$ for several values of the average node degree  $\langle{k}\rangle$ and two values of the threshold: (a) $r = 0.5$ and (b) $r = 0.6$. Thin (red and blue colors online) lines refer to different values of $\langle{k}\rangle \in \lbrace{10,20,40,80}\rbrace$ from left to right, whereas thick black lines represent the mean-field solution. Arrows indicate the direction in which $\langle{k}\rangle$ increases.}
	\label{fig:c_PA}
\end{figure}

Now it is time to investigate the second ingredient, namely to check whether $\langle{k}\rangle$ influences phase transitions within ST model. In Fig. \ref{fig:c_PA} we present the dependence between the stationary concentration of spins up $c$ and the noise $p$ for several values of the average node degree of the network $\langle{k}\rangle$ and two values of the threshold $r$. Again we see that for $r=0.5$ only continuous phase transitions are observed independently on $\langle{k}\rangle$. However, for $r=0.6$ the character of the phase transition changes with $\langle{k}\rangle$. Similarly as for the MV model with inertia and the qV  model, the width of the hysteresis increases with  $\langle{k}\rangle$ \cite{Che:etal:17,Enc:etal:18,Abr:Szn:20}.

Due to our knowledge, the dependence between the size of the hysteresis and $\langle{k}\rangle$  was not investigated precisely for the MV model with inertia. However, for the $q$-voter model it has been shown that $\langle{k}\rangle$ influences substantially the width of the hysteresis and has almost no influence to the jump of the order parameter, defined as \cite{Abr:Szn:20}:
\begin{equation}
m=\frac{N_{\uparrow} - N_{\downarrow}}{N}=2\frac{N_{\uparrow}}{N}-1=2c-1.
\label{eq:mag}
\end{equation}
In this paper we did not introduce order parameter $m$, because we made all calculations in terms of $c$. Of course we could easily reformulate all results using the simple relation between $m$ and $c$, given by Eq. (\ref{eq:mag}).

\begin{figure}[t!]
	\includegraphics[width=0.48\textwidth]{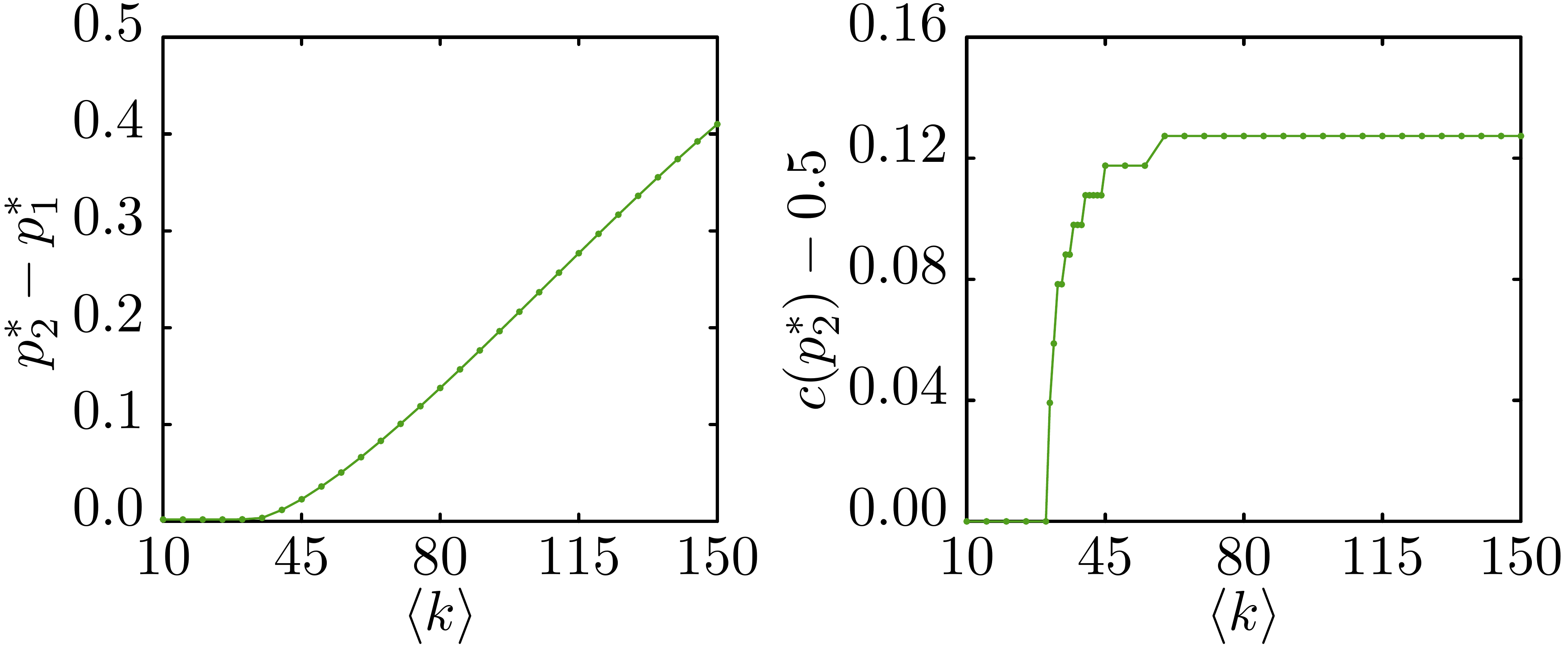}
	\caption{The width of hysteresis $p_2^*-p_1^*$ (left panel) and the jump of the public opinion $c$ (right panel) as a function of the average node degree $\langle{k}\rangle$ for threshold $r = 0.6$ obtained within PA.}
	\label{fig:dep_k}
\end{figure}

In \cite{Abr:Szn:20} the jump of $m$ has been measured at upper spinodal. Therefore we also measure a jump of $c$ at this point, i.e. $c(p_2^*)-0.5$. As we see in Fig. \ref{fig:dep_k} both hysteresis, as well as the jump of $c$ depend on $\langle{k}\rangle$. However, these dependencies are very different. There is only one common feature seen in both relations -- below certain value of $\langle{k}\rangle$ both $p_2^*-p_1^*$, as well as $c(p_2^*)-0.5$ are equal zero, which indicates continuous phase transition. Above this value the width of hysteresis increases with $\langle{k}\rangle$ almost linearly. On the other hand the jump of concentration of spins up increases only slightly but this growth is very rapid and takes place in a relatively small range of $\langle{k}\rangle$. For larger values of $\langle{k}\rangle$  the jump of $c$ does not change, similarly as for the $q$-voter model \cite{Abr:Szn:20}. 

\begin{figure}[b]
	\includegraphics[width=0.47\textwidth]{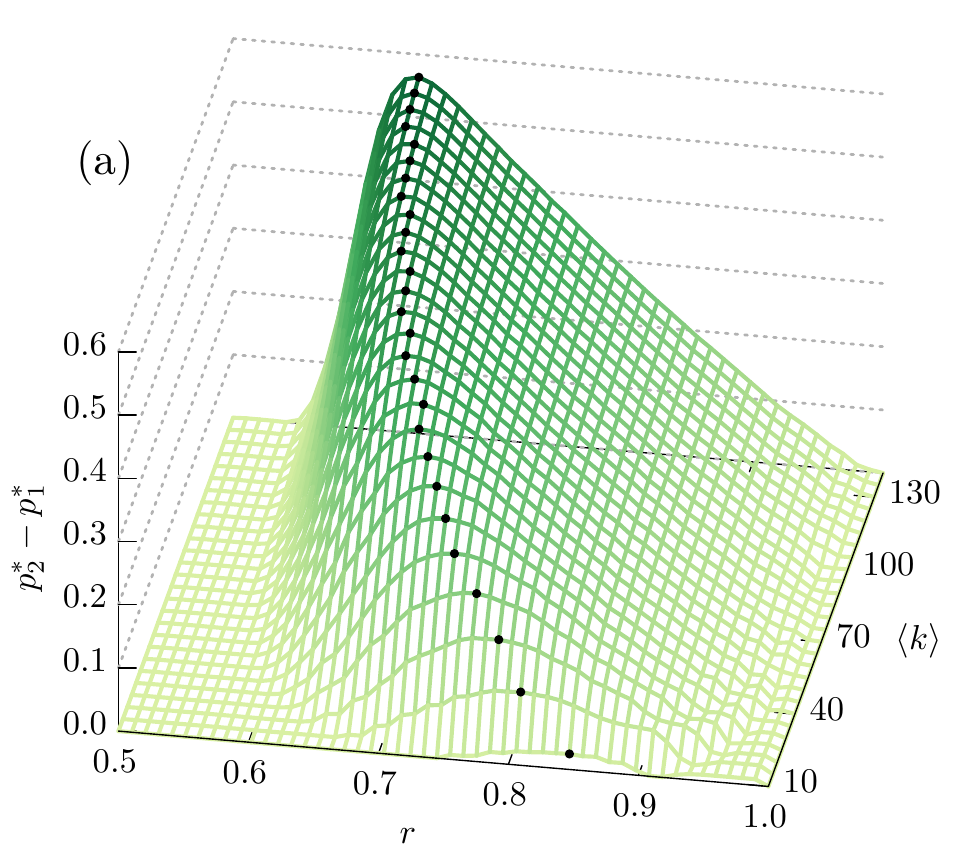}
	\includegraphics[width=0.47\textwidth]{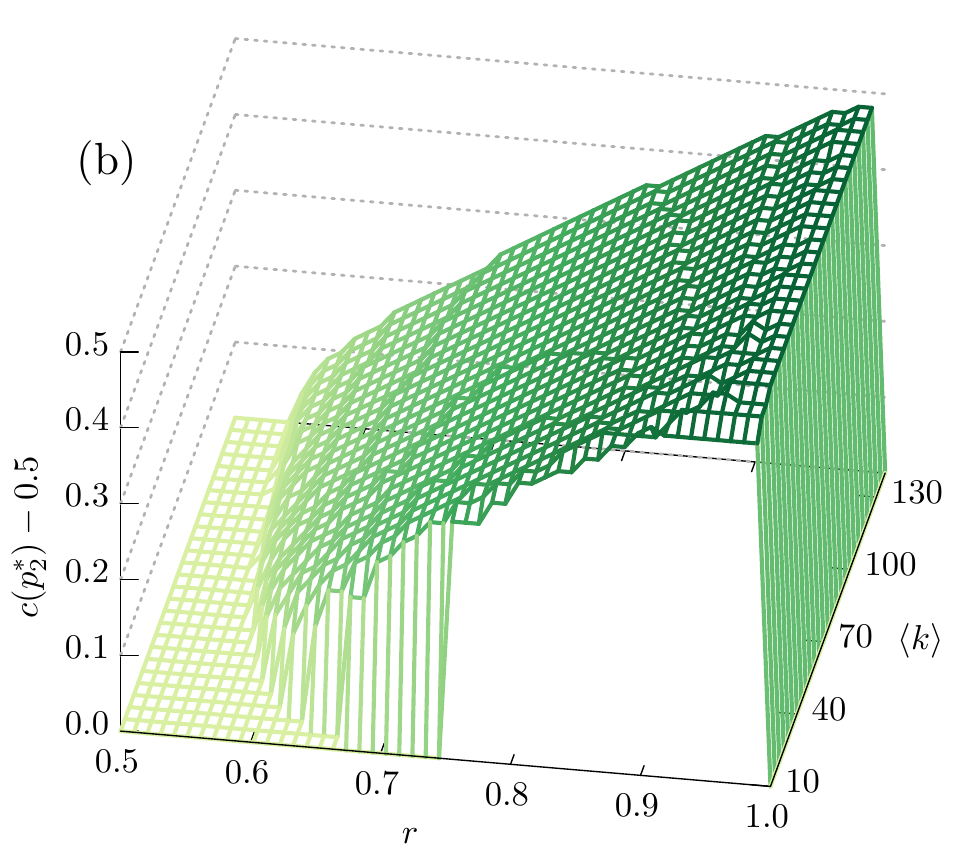}
	\caption{The size of the hysteresis (a) and the jump of the concentration $c$ at upper spinodal $p_{2}^{*}$ (b) as a function of the threshold $r$ and the average degree of a graph $\langle{k}\rangle$ obtained within PA.} 
	\label{fig:3d}
\end{figure}

Until now we have analyzed the influence of $\langle{k}\rangle$ on the phase transition only for $r=0.6$. Of course the same can be done for an arbitrary value of $r$, as shown in Fig. \ref{fig:3d}. We see that the width of the hysteresis indeed increases monotonically with $\langle{k}\rangle$. However, the dependence on the threshold $r$ is much more interesting. There is an optimal value of $r$, which decreases with $\langle{k}\rangle$, for which the hysteresis has the maximum size. Because empirical studies suggest that the mean number of friends varies typically from $5$ to $150$, depending on the rated emotional closeness between them, \cite{Dun:etal:15}, optimal value of $r$, for which the maximum size of hysteresis appear lies in $(0.65,0.85)$. We find this result particularly interesting from the social point of view, which will be commented in the Conclusions.

\section{\label{sec:MC_simulations}Monte Carlo Simulations}
We validate our analytical PA results by Monte Carlo (MC) simulations on WS graphs \cite{Watt:Str:98}. As we have written in the introduction, WS algorithm allows to tune the structure of the graph from a regular ($\beta = 0$) to a random one ($\beta = 1$). It also reduces to the complete graph for $\langle{k}\rangle=N-1$. Moreover, in the whole spectrum of parameter $\beta$ the average node degree is conserved. This makes the WS graph particularly interesting for our studies.

\begin{figure*}[t]
	\includegraphics[width=0.9\textwidth]{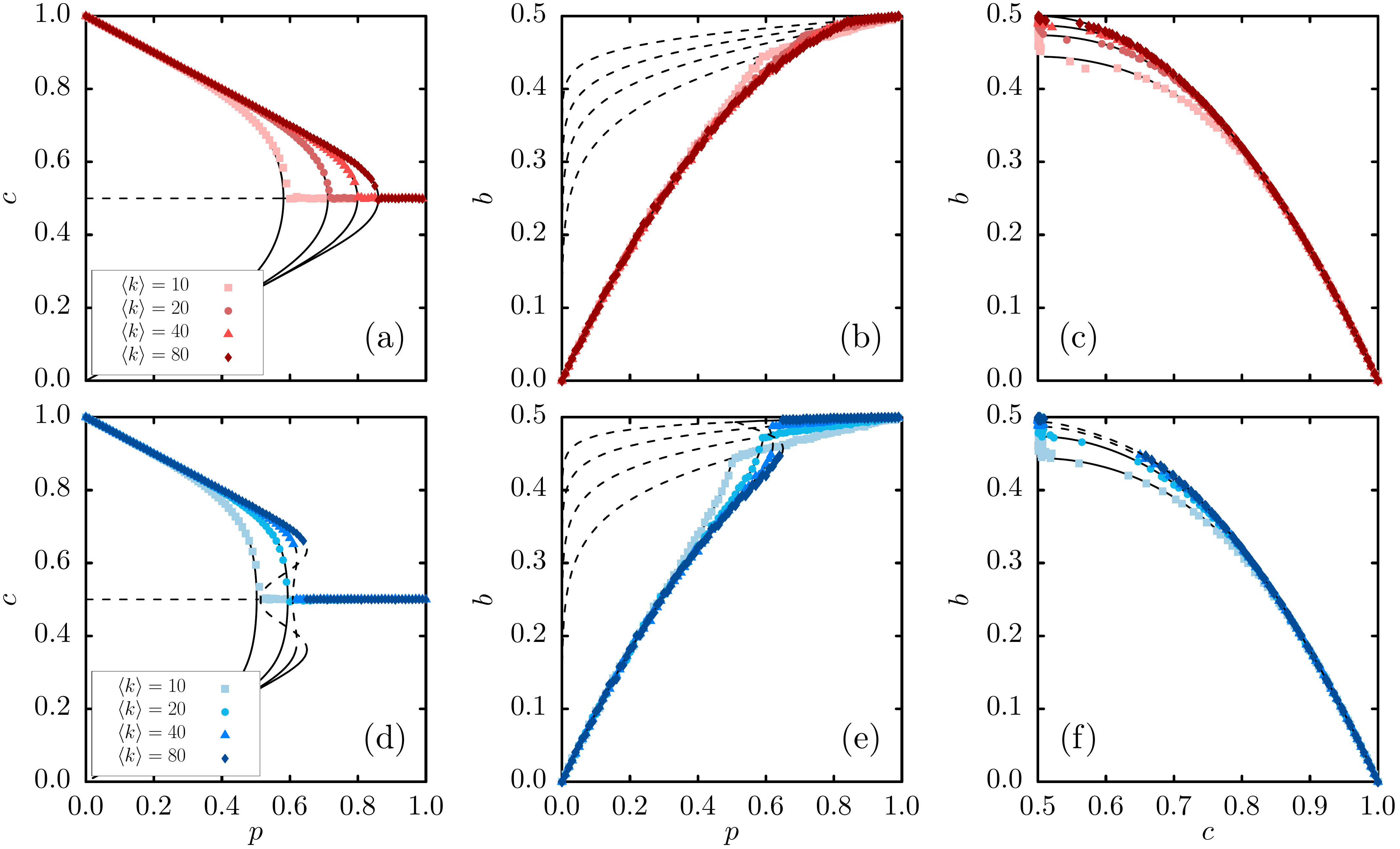}
	\caption{Comparison between results obtained within PA (denoted by lines) and Monte Carlo simulations (denoted by symbols) for $r=0.5$ (upper panels) and $r=0.6$ (bottom panels). Solid lines correspond to stable, whereas dashed lines to unstable solutions of Eqs. (\ref{eq:stat:c})--(\ref{eq:stat:b}). For all diagrams the size of the system $N = 10^4$, the thermalization time $t=10^4$ and the initial concentration of spins up $c(0)=1$. Results are averaged only over $5$ samples, but for this size of the system it is sufficient, as seen above.}
	\label{fig:MC_PA}
\end{figure*}

We start with $\beta = 1$, for which PA should be the most accurate. Indeed, as seen in Fig. \ref{fig:MC_PA}, Monte Carlo overlap PA results, even for small values of $\langle{k}\rangle$. Moreover, this agreement is seen in all dependencies, namely $c=c(p),b=b(p),b=b(c)$. The question is if and how parameter $\beta$ will influence results. 

\begin{figure*}[t]
	\includegraphics[width=0.9\textwidth]{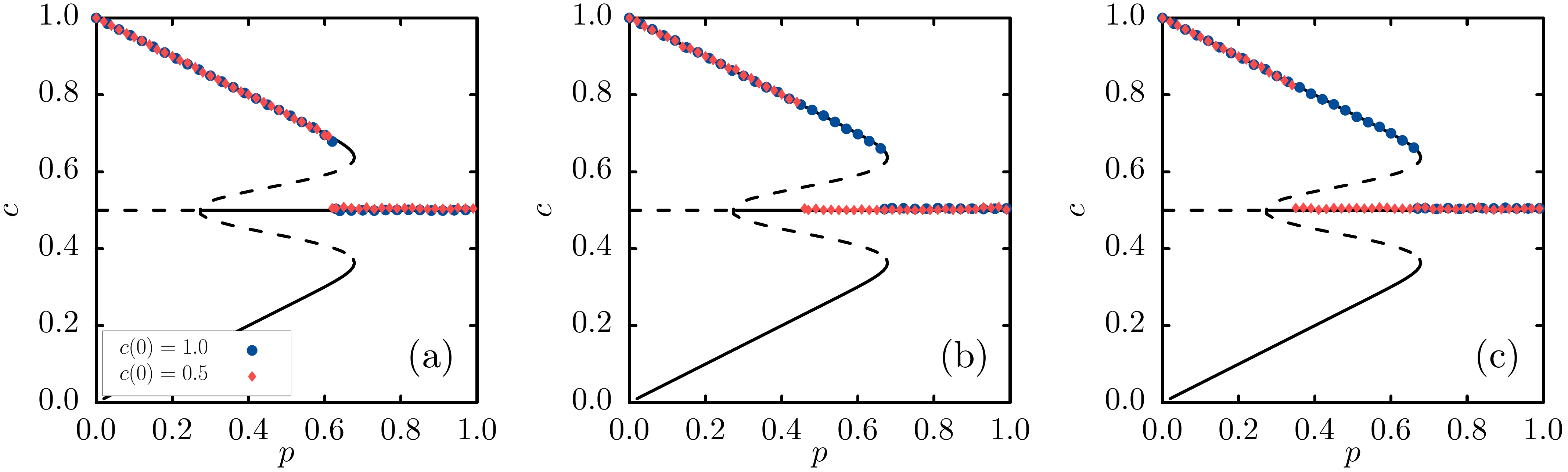}
	\caption{Dependence between the stationary concentration of spins up $c$ and the noise $p$ for $r=0.6$, $\langle{k}\rangle = 150$,  and several values of rewriting parameter: (a) $\beta = 0.1$, (b) $\beta = 0.5$, (c) $\beta = 1$. Monte Carlo results for two types of initial conditions and $N=10^4$ are denoted by symbols, whereas lines correspond to PA results. As in Fig.\ref{fig:MC_PA}, thermalization time $t=10^4$ and results are averaged over $5$ samples.}
	\label{fig:cp:beta}
\end{figure*}

In Fig. \ref{fig:cp:beta} parameter $\beta$ vary from $0.1$ to $1$. As seen, the width of the hysteresis $p^*_2 - p^*_1$ is increasing with $\beta$. Such a tendency is seen for all values of $r$. As usually, in general PA gives consistent results with MC simulations only for sufficiently large values of the rewiring probability $\beta$. However, as seen in Fig. \ref{fig:cp:beta}, the value of the upper spinodal is less sensitive to $\beta$ than the lower spinodal and is predicted correctly even for $\beta = 0.5$. 

\section{\label{sec:Conclusions}Conclusions}
The notion of the tipping point, similarly as the notion of the hysteresis, two signatures of  discontinuous phase transitions, has been present in social sciences for many years \cite{Sch:Wes:Bro:03}. Although it may seem that the social hysteresis and the tipping point are just fancy buzzwords, empirical social studies have confirmed that they are not just abstract ideas \cite{Sch:Wes:Bro:03,Doe:etal:18,Cen:etal:18}. 

These findings, among others, inspired researchers to look for the hysteresis in models of opinion dynamics \cite{Che:etal:17,Enc:etal:18,Abr:Szn:20}. For example, an additional noise has been introduce to the MV model, but is was shown that it does not affect the type of the phase transition and it remains continuous irrespective of the network degree and its distribution \cite{Vie:Cro:16,Enc:etal:19}. On the other hand it was shown that discontinuous phase transitions may appear in the MV model with inertia, when the inertia is above an appropriate level \cite{Che:etal:17}. Later the question about the \textit{fundamental ingredients for discontinuous phase transitions in the inertial majority vote model} has been asked \cite{Enc:etal:18}. It was shown that low $\langle{k}\rangle$  leads to the suppression of the phase coexistence. Similar result has been also reported for the $q$-voter model \cite{Abr:Szn:20}.

This motivated us to check if the same behavior will be observed within the symmetrical threshold model introduced in \cite{Now:Szn:19}. We have shown, using PA and MC simulations, that indeed the type of the phase transition within ST model depends on threshold $r$, as well as the properties of the network  $\langle{k}\rangle$ and $\beta$, i.e. hysteresis increases with $\langle{k}\rangle$ and $\beta$. On the other hand, the dependence on $r$ is non-monotonic, which will be commented below.

We discuss ST in the context of MV and qV models, because they have a lot in common, which has been already discussed in \cite{Now:Szn:19}. In particular, ST model with anticonformity is the generalization of the basic majority-vote model, which corresponds to $r=0.5$. Moreover, ST model with $r=1$ reduces to the $q$-voter model on the random regular graph with degree $q$, i.e. if $\forall i k_i=k=q$. Finally, ST model with an arbitrary value of $r$ corresponds to the threshold $q$-voter model on the random regular graph with $\forall i k_i=k=q$ \cite{Nyc:Szn:13,Nyc:etal:18,Vie:Ant:18,Vie:etal:20}. 

Moreover, as we have noticed in Sec. \ref{sec:discussion}, the parameters that are mainly responsible for the discontinuous phase transitions, namely: the level of inertia $\theta$ in the MV model with inertia, the size of the influence group $q$ in the qV model and the threshold $r$ needed for the social influence in the ST model, play in a sense a similar role. The larger $q$ or $r$ is, the harder it is to influence a voter, which in result increases inertia on the microscopic level.

Because the hysteresis can be viewed as an inertia of the system on the macroscopic level, it would not be surprising that the inertia on the microscopic level supports the hysteresis. However, as shown in Fig. \ref{fig:3d}, the relation between the size of the hysteresis and parameter $r$ is not that trivial, i.e. it is non-monotonic, having the maximum value for a given value of $r$, which depends on $\langle{k}\rangle$. This is particularly interesting result from the social point view and worth to be discussed here.

It is known that social influence increases with the size of the influence group as well as the unanimity of the group. However, this dependence is far from being trivial. First of all, it occurs that it increases only up to a certain level. The social influence is stronger if the group of influence consists of $4$, instead of $2$ people. However, above a certain threshold it remains on the same level. Moreover, above this threshold, around $7-11$ people, the social influence decreases \cite{Asch:55}. 

Therefore, in social experiments, in which descriptive norms are used to influence people, social psychologists neither use unanimity nor simple majority. Instead they use certain super-majority, often around $75\%$. For example they manipulate people to reuse towels in hotels with the fake descriptive norm saying something like: $75\%$ of our guests are reusing towels". There is no strong evidence that $75\%$ is the magic number and in some other experiments larger majorities were used as briefly reviewed in \cite{Nyc:etal:18}. The main message we want to pass here is that the larger majority does not always result in stronger social influence. It seems that some optimal values exist and these values probably depend on the size of the influence group: for small groups unanimity is needed but for large groups some threshold value is more appropriate, significantly larger than $50\%$, but smaller than $100\%$. How this is related with the results obtained here?

As we have already written in Section \ref{sec:discussion}, it was found empirically that in real social networks $\langle{k}\rangle \in (5,150)$. For these values the optimal threshold of $r$, for which the largest social hysteresis is observed, lies in the range $(0.65,0.85)$, depending on the average size of the influence group $\langle{k}\rangle$. 
We admit that what we measure is not the power of social influence, but the size of the hysteresis. However, having in mind that the hysteresis is usually observed in social systems, we can speculate that  there are some optimal values in the level of social influence and these values influence the hysteresis, which is usually observed in social systems.

We are aware that it maybe merely intriguing but the meaningless coincident. However, we believe that this finding deserves more attention and studies within other models of opinion dynamics.

\section*{Acknowledgments}
This work is supported by funds from the National Science Centre (NCN, Poland) through grant
no. 2016/21/B/HS6/01256 and by PLGrid Infrastructure. 

\section*{Data Availability Statement}
Data sharing is not applicable to this article as no new data were created or analyzed in this study.

%


\begin{thebibliography}{46}%
	\makeatletter
	\providecommand \@ifxundefined [1]{%
		\@ifx{#1\undefined}
	}%
	\providecommand \@ifnum [1]{%
		\ifnum #1\expandafter \@firstoftwo
		\else \expandafter \@secondoftwo
		\fi
	}%
	\providecommand \@ifx [1]{%
		\ifx #1\expandafter \@firstoftwo
		\else \expandafter \@secondoftwo
		\fi
	}%
	\providecommand \natexlab [1]{#1}%
	\providecommand \enquote  [1]{``#1''}%
	\providecommand \bibnamefont  [1]{#1}%
	\providecommand \bibfnamefont [1]{#1}%
	\providecommand \citenamefont [1]{#1}%
	\providecommand \href@noop [0]{\@secondoftwo}%
	\providecommand \href [0]{\begingroup \@sanitize@url \@href}%
	\providecommand \@href[1]{\@@startlink{#1}\@@href}%
	\providecommand \@@href[1]{\endgroup#1\@@endlink}%
	\providecommand \@sanitize@url [0]{\catcode `\\12\catcode `\$12\catcode
		`\&12\catcode `\#12\catcode `\^12\catcode `\_12\catcode `\%12\relax}%
	\providecommand \@@startlink[1]{}%
	\providecommand \@@endlink[0]{}%
	\providecommand \url  [0]{\begingroup\@sanitize@url \@url }%
	\providecommand \@url [1]{\endgroup\@href {#1}{\urlprefix }}%
	\providecommand \urlprefix  [0]{URL }%
	\providecommand \Eprint [0]{\href }%
	\providecommand \doibase [0]{http://dx.doi.org/}%
	\providecommand \selectlanguage [0]{\@gobble}%
	\providecommand \bibinfo  [0]{\@secondoftwo}%
	\providecommand \bibfield  [0]{\@secondoftwo}%
	\providecommand \translation [1]{[#1]}%
	\providecommand \BibitemOpen [0]{}%
	\providecommand \bibitemStop [0]{}%
	\providecommand \bibitemNoStop [0]{.\EOS\space}%
	\providecommand \EOS [0]{\spacefactor3000\relax}%
	\providecommand \BibitemShut  [1]{\csname bibitem#1\endcsname}%
	\let\auto@bib@innerbib\@empty
	\bibitem [{\citenamefont {Watts}\ and\ \citenamefont
		{Dodds}(2017)}]{Wat:Dod:17}%
	\BibitemOpen
	\bibfield  {author} {\bibinfo {author} {\bibfnamefont {D.}~\bibnamefont
			{Watts}}\ and\ \bibinfo {author} {\bibfnamefont {P.}~\bibnamefont {Dodds}},\
	}\href {\doibase 10.1093/oxfordhb/9780199215362.013.20} {\emph {\bibinfo
			{title} {Threshold models of social influence}}}\ (\bibinfo {year} {2017})\
	pp.\ \bibinfo {pages} {475--497}\BibitemShut {NoStop}%
	\bibitem [{\citenamefont {J\k{e}drzejewski}\ and\ \citenamefont
		{Sznajd-Weron}(2019)}]{Jed:Szn:19}%
	\BibitemOpen
	\bibfield  {author} {\bibinfo {author} {\bibfnamefont {A.}~\bibnamefont
			{J\k{e}drzejewski}}\ and\ \bibinfo {author} {\bibfnamefont {K.}~\bibnamefont
			{Sznajd-Weron}},\ }\href {\doibase 10.1016/j.crhy.2019.05.002} {\bibfield
		{journal} {\bibinfo  {journal} {Comptes Rendus Physique}\ } (\bibinfo {year}
		{2019}),\ 10.1016/j.crhy.2019.05.002}\BibitemShut {NoStop}%
	\bibitem [{\citenamefont {Watts}(2002)}]{Wat:02}%
	\BibitemOpen
	\bibfield  {author} {\bibinfo {author} {\bibfnamefont {D.~J.}\ \bibnamefont
			{Watts}},\ }\href {\doibase 10.1073/pnas.082090499} {\bibfield  {journal}
		{\bibinfo  {journal} {Proceedings of the National Academy of Sciences of the
				United States of America}\ }\textbf {\bibinfo {volume} {99}},\ \bibinfo
		{pages} {5766} (\bibinfo {year} {2002})}\BibitemShut {NoStop}%
	\bibitem [{\citenamefont {Grabisch}\ and\ \citenamefont
		{Li}(2019)}]{Gra:Li:19}%
	\BibitemOpen
	\bibfield  {author} {\bibinfo {author} {\bibfnamefont {M.}~\bibnamefont
			{Grabisch}}\ and\ \bibinfo {author} {\bibfnamefont {F.}~\bibnamefont {Li}},\
	}\href {\doibase 10.1007/s13235-019-00332-0} {\bibfield  {journal} {\bibinfo
			{journal} {Dynamic Games and Applications}\ } (\bibinfo {year} {2019}),\
		10.1007/s13235-019-00332-0}\BibitemShut {NoStop}%
	\bibitem [{\citenamefont {Granovetter}(1978)}]{Gra:78}%
	\BibitemOpen
	\bibfield  {author} {\bibinfo {author} {\bibfnamefont {M.}~\bibnamefont
			{Granovetter}},\ }\href {\doibase 10.1086/226707} {\bibfield  {journal}
		{\bibinfo  {journal} {Am. J. Sociol.}\ }\textbf {\bibinfo {volume} {83}}
		(\bibinfo {year} {1978}),\ 10.1086/226707}\BibitemShut {NoStop}%
	\bibitem [{\citenamefont {Galam}(1990)}]{Gal:90}%
	\BibitemOpen
	\bibfield  {author} {\bibinfo {author} {\bibfnamefont {S.}~\bibnamefont
			{Galam}},\ }\href {\doibase 10.1007/BF01027314} {\bibfield  {journal}
		{\bibinfo  {journal} {J. Stat. Phys.}\ }\textbf {\bibinfo {volume} {61}},\
		\bibinfo {pages} {943} (\bibinfo {year} {1990})}\BibitemShut {NoStop}%
	\bibitem [{\citenamefont {Galam}(2008)}]{Gal:08}%
	\BibitemOpen
	\bibfield  {author} {\bibinfo {author} {\bibfnamefont {S.}~\bibnamefont
			{Galam}},\ }\href {\doibase 10.1142/S0129183108012297} {\bibfield  {journal}
		{\bibinfo  {journal} {International Journal of Modern Physics C}\ }\textbf
		{\bibinfo {volume} {19}},\ \bibinfo {pages} {409} (\bibinfo {year}
		{2008})}\BibitemShut {NoStop}%
	\bibitem [{\citenamefont {Galam}(2012)}]{Gal:12}%
	\BibitemOpen
	\bibfield  {author} {\bibinfo {author} {\bibfnamefont {S.}~\bibnamefont
			{Galam}},\ }\href {\doibase 10.1007/978-1-4614-2032-3} {\emph {\bibinfo
			{title} {Sociophysics: A Physicist's Modeling of Psycho-Political
				Phenomena}}}\ (\bibinfo  {publisher} {Springer-Verlag},\ \bibinfo {year}
	{2012})\BibitemShut {NoStop}%
	\bibitem [{\citenamefont {Liggett}(1985)}]{Lig:85}%
	\BibitemOpen
	\bibfield  {author} {\bibinfo {author} {\bibfnamefont {T.~M.}\ \bibnamefont
			{Liggett}},\ }\href {\doibase 10.1007/b138374} {\emph {\bibinfo {title}
			{Interacting Particle Systems}}}\ (\bibinfo  {publisher} {Springer},\
	\bibinfo {year} {1985})\BibitemShut {NoStop}%
	\bibitem [{\citenamefont {Tome}\ \emph {et~al.}(1991)\citenamefont {Tome},
		\citenamefont {De~Oliveira},\ and\ \citenamefont {Santos}}]{Tom:Oli:San:91}%
	\BibitemOpen
	\bibfield  {author} {\bibinfo {author} {\bibfnamefont {T.}~\bibnamefont
			{Tome}}, \bibinfo {author} {\bibfnamefont {M.}~\bibnamefont {De~Oliveira}}, \
		and\ \bibinfo {author} {\bibfnamefont {M.}~\bibnamefont {Santos}},\ }\href
	{\doibase 10.1088/0305-4470/24/15/033} {\bibfield  {journal} {\bibinfo
			{journal} {Journal of Physics A: General Physics}\ }\textbf {\bibinfo
			{volume} {24}},\ \bibinfo {pages} {3677} (\bibinfo {year}
		{1991})}\BibitemShut {NoStop}%
	\bibitem [{\citenamefont {de~Oliveira}(1992)}]{Oli:92}%
	\BibitemOpen
	\bibfield  {author} {\bibinfo {author} {\bibfnamefont {M.}~\bibnamefont
			{de~Oliveira}},\ }\href {\doibase 10.1007/BF01060069} {\bibfield  {journal}
		{\bibinfo  {journal} {Journal of Statistical Physics}\ }\textbf {\bibinfo
			{volume} {66}},\ \bibinfo {pages} {273} (\bibinfo {year} {1992})}\BibitemShut
	{NoStop}%
	\bibitem [{\citenamefont {Lima}\ and\ \citenamefont
		{Malarz}(2006)}]{Lim:Mal:06}%
	\BibitemOpen
	\bibfield  {author} {\bibinfo {author} {\bibfnamefont {F.}~\bibnamefont
			{Lima}}\ and\ \bibinfo {author} {\bibfnamefont {K.}~\bibnamefont {Malarz}},\
	}\href {\doibase 10.1142/S0129183106009849} {\bibfield  {journal} {\bibinfo
			{journal} {International Journal of Modern Physics C}\ }\textbf {\bibinfo
			{volume} {17}},\ \bibinfo {pages} {1273} (\bibinfo {year}
		{2006})}\BibitemShut {NoStop}%
	\bibitem [{\citenamefont {Santos}\ \emph {et~al.}(2010)\citenamefont {Santos},
		\citenamefont {Lima},\ and\ \citenamefont {Malarz}}]{San:Lim:Mal:10}%
	\BibitemOpen
	\bibfield  {author} {\bibinfo {author} {\bibfnamefont {J.}~\bibnamefont
			{Santos}}, \bibinfo {author} {\bibfnamefont {F.}~\bibnamefont {Lima}}, \ and\
		\bibinfo {author} {\bibfnamefont {K.}~\bibnamefont {Malarz}},\ }\href
	{\doibase 10.1016/j.physa.2010.08.054} {\bibfield  {journal} {\bibinfo
			{journal} {Physica A: Statistical Mechanics and its Applications}\ }\textbf
		{\bibinfo {volume} {390}},\ \bibinfo {pages} {359} (\bibinfo {year}
		{2010})}\BibitemShut {NoStop}%
	\bibitem [{\citenamefont {Vieira}\ and\ \citenamefont
		{Crokidakis}(2016)}]{Vie:Cro:16}%
	\BibitemOpen
	\bibfield  {author} {\bibinfo {author} {\bibfnamefont {A.}~\bibnamefont
			{Vieira}}\ and\ \bibinfo {author} {\bibfnamefont {N.}~\bibnamefont
			{Crokidakis}},\ }\href {\doibase 10.1016/j.physa.2016.01.013} {\bibfield
		{journal} {\bibinfo  {journal} {Physica A: Statistical Mechanics and its
				Applications}\ }\textbf {\bibinfo {volume} {450}},\ \bibinfo {pages} {30}
		(\bibinfo {year} {2016})}\BibitemShut {NoStop}%
	\bibitem [{\citenamefont {Chen}\ \emph {et~al.}(2017)\citenamefont {Chen},
		\citenamefont {Shen}, \citenamefont {Zhang}, \citenamefont {Li},
		\citenamefont {Hou},\ and\ \citenamefont {Kurths}}]{Che:etal:17}%
	\BibitemOpen
	\bibfield  {author} {\bibinfo {author} {\bibfnamefont {H.}~\bibnamefont
			{Chen}}, \bibinfo {author} {\bibfnamefont {C.}~\bibnamefont {Shen}}, \bibinfo
		{author} {\bibfnamefont {H.}~\bibnamefont {Zhang}}, \bibinfo {author}
		{\bibfnamefont {G.}~\bibnamefont {Li}}, \bibinfo {author} {\bibfnamefont
			{Z.}~\bibnamefont {Hou}}, \ and\ \bibinfo {author} {\bibfnamefont
			{J.}~\bibnamefont {Kurths}},\ }\href {\doibase 10.1103/PhysRevE.95.042304}
	{\bibfield  {journal} {\bibinfo  {journal} {Physical Review E}\ }\textbf
		{\bibinfo {volume} {95}} (\bibinfo {year} {2017}),\
		10.1103/PhysRevE.95.042304}\BibitemShut {NoStop}%
	\bibitem [{\citenamefont {Fronczak}\ and\ \citenamefont
		{Fronczak}(2017)}]{Fro:Fro:17}%
	\BibitemOpen
	\bibfield  {author} {\bibinfo {author} {\bibfnamefont {A.}~\bibnamefont
			{Fronczak}}\ and\ \bibinfo {author} {\bibfnamefont {P.}~\bibnamefont
			{Fronczak}},\ }\href {\doibase 10.1103/PhysRevE.96.012304} {\bibfield
		{journal} {\bibinfo  {journal} {Physical Review E}\ }\textbf {\bibinfo
			{volume} {96}} (\bibinfo {year} {2017}),\
		10.1103/PhysRevE.96.012304}\BibitemShut {NoStop}%
	\bibitem [{\citenamefont {Krawiecki}(2018)}]{Kra:18}%
	\BibitemOpen
	\bibfield  {author} {\bibinfo {author} {\bibfnamefont {A.}~\bibnamefont
			{Krawiecki}},\ }\href {\doibase 10.1140/epjb/e2018-80551-9} {\bibfield
		{journal} {\bibinfo  {journal} {European Physical Journal B}\ }\textbf
		{\bibinfo {volume} {91}} (\bibinfo {year} {2018}),\
		10.1140/epjb/e2018-80551-9}\BibitemShut {NoStop}%
	\bibitem [{\citenamefont {Krawiecki}\ and\ \citenamefont
		{Gradowski}(2019)}]{Kra:Gra:19}%
	\BibitemOpen
	\bibfield  {author} {\bibinfo {author} {\bibfnamefont {A.}~\bibnamefont
			{Krawiecki}}\ and\ \bibinfo {author} {\bibfnamefont {T.}~\bibnamefont
			{Gradowski}},\ }\href {\doibase 10.5506/APhysPolBSupp.12.91} {\bibfield
		{journal} {\bibinfo  {journal} {Acta Physica Polonica B, Proceedings
				Supplement}\ }\textbf {\bibinfo {volume} {12}},\ \bibinfo {pages} {91}
		(\bibinfo {year} {2019})}\BibitemShut {NoStop}%
	\bibitem [{\citenamefont {Encinas}\ \emph {et~al.}(2018)\citenamefont
		{Encinas}, \citenamefont {Harunari}, \citenamefont {De~Oliveira},\ and\
		\citenamefont {Fiore}}]{Enc:etal:18}%
	\BibitemOpen
	\bibfield  {author} {\bibinfo {author} {\bibfnamefont {J.}~\bibnamefont
			{Encinas}}, \bibinfo {author} {\bibfnamefont {P.}~\bibnamefont {Harunari}},
		\bibinfo {author} {\bibfnamefont {M.}~\bibnamefont {De~Oliveira}}, \ and\
		\bibinfo {author} {\bibfnamefont {C.}~\bibnamefont {Fiore}},\ }\href
	{\doibase 10.1038/s41598-018-27240-4} {\bibfield  {journal} {\bibinfo
			{journal} {Scientific Reports}\ }\textbf {\bibinfo {volume} {8}} (\bibinfo
		{year} {2018}),\ 10.1038/s41598-018-27240-4}\BibitemShut {NoStop}%
	\bibitem [{\citenamefont {Encinas}\ \emph {et~al.}(2019)\citenamefont
		{Encinas}, \citenamefont {Chen}, \citenamefont {de~Oliveira},\ and\
		\citenamefont {Fiore}}]{Enc:etal:19}%
	\BibitemOpen
	\bibfield  {author} {\bibinfo {author} {\bibfnamefont {J.}~\bibnamefont
			{Encinas}}, \bibinfo {author} {\bibfnamefont {H.}~\bibnamefont {Chen}},
		\bibinfo {author} {\bibfnamefont {M.}~\bibnamefont {de~Oliveira}}, \ and\
		\bibinfo {author} {\bibfnamefont {C.}~\bibnamefont {Fiore}},\ }\href
	{\doibase 10.1016/j.physa.2018.10.055} {\bibfield  {journal} {\bibinfo
			{journal} {Physica A: Statistical Mechanics and its Applications}\ }\textbf
		{\bibinfo {volume} {516}},\ \bibinfo {pages} {563} (\bibinfo {year}
		{2019})}\BibitemShut {NoStop}%
	\bibitem [{\citenamefont {Castellano}\ \emph {et~al.}(2009)\citenamefont
		{Castellano}, \citenamefont {Mu{\~n}oz},\ and\ \citenamefont
		{Pastor-Satorras}}]{Cas:Mun:Pat:09}%
	\BibitemOpen
	\bibfield  {author} {\bibinfo {author} {\bibfnamefont {C.}~\bibnamefont
			{Castellano}}, \bibinfo {author} {\bibfnamefont {M.~A.}\ \bibnamefont
			{Mu{\~n}oz}}, \ and\ \bibinfo {author} {\bibfnamefont {R.}~\bibnamefont
			{Pastor-Satorras}},\ }\href {\doibase 10.1103/PhysRevE.80.041129} {\bibfield
		{journal} {\bibinfo  {journal} {Physical Review E}\ }\textbf {\bibinfo
			{volume} {80}},\ \bibinfo {pages} {041129} (\bibinfo {year}
		{2009})}\BibitemShut {NoStop}%
	\bibitem [{\citenamefont {Nyczka}\ \emph
		{et~al.}(2012{\natexlab{a}})\citenamefont {Nyczka}, \citenamefont
		{Sznajd-Weron},\ and\ \citenamefont {Cis\l{}o}}]{Nyc:Szn:Cis:12}%
	\BibitemOpen
	\bibfield  {author} {\bibinfo {author} {\bibfnamefont {P.}~\bibnamefont
			{Nyczka}}, \bibinfo {author} {\bibfnamefont {K.}~\bibnamefont
			{Sznajd-Weron}}, \ and\ \bibinfo {author} {\bibfnamefont {J.}~\bibnamefont
			{Cis\l{}o}},\ }\href {\doibase 10.1103/PhysRevE.86.011105} {\bibfield
		{journal} {\bibinfo  {journal} {Phys. Rev. E}\ }\textbf {\bibinfo {volume}
			{86}},\ \bibinfo {pages} {011105} (\bibinfo {year}
		{2012}{\natexlab{a}})}\BibitemShut {NoStop}%
	\bibitem [{\citenamefont {Moretti}\ \emph {et~al.}(2013)\citenamefont
		{Moretti}, \citenamefont {Liu}, \citenamefont {Castellano},\ and\
		\citenamefont {Pastor-Satorras}}]{Mor:etal:13}%
	\BibitemOpen
	\bibfield  {author} {\bibinfo {author} {\bibfnamefont {P.}~\bibnamefont
			{Moretti}}, \bibinfo {author} {\bibfnamefont {S.}~\bibnamefont {Liu}},
		\bibinfo {author} {\bibfnamefont {C.}~\bibnamefont {Castellano}}, \ and\
		\bibinfo {author} {\bibfnamefont {R.}~\bibnamefont {Pastor-Satorras}},\
	}\href {\doibase 10.1007/s10955-013-0704-1} {\bibfield  {journal} {\bibinfo
			{journal} {Journal of Statistical Physics}\ }\textbf {\bibinfo {volume}
			{151}},\ \bibinfo {pages} {113} (\bibinfo {year} {2013})}\BibitemShut
	{NoStop}%
	\bibitem [{\citenamefont {Mobilia}(2015)}]{Mob:15}%
	\BibitemOpen
	\bibfield  {author} {\bibinfo {author} {\bibfnamefont {M.}~\bibnamefont
			{Mobilia}},\ }\href {\doibase 10.1103/PhysRevE.92.012803} {\bibfield
		{journal} {\bibinfo  {journal} {Physical Review E}\ }\textbf {\bibinfo
			{volume} {92}},\ \bibinfo {pages} {012803} (\bibinfo {year}
		{2015})}\BibitemShut {NoStop}%
	\bibitem [{\citenamefont {Javarone}\ and\ \citenamefont
		{Squartini}(2015)}]{Jav:Squ:15}%
	\BibitemOpen
	\bibfield  {author} {\bibinfo {author} {\bibfnamefont {M.~A.}\ \bibnamefont
			{Javarone}}\ and\ \bibinfo {author} {\bibfnamefont {T.}~\bibnamefont
			{Squartini}},\ }\href {\doibase 10.1088/1742-5468/2015/10/P10002} {\bibfield
		{journal} {\bibinfo  {journal} {Journal of Statistical Mechanics: Theory and
				Experiment}\ }\textbf {\bibinfo {volume} {2015}},\ \bibinfo {pages} {P10002}
		(\bibinfo {year} {2015})}\BibitemShut {NoStop}%
	\bibitem [{\citenamefont {Mellor}\ \emph {et~al.}(2016)\citenamefont {Mellor},
		\citenamefont {Mobilia},\ and\ \citenamefont {Zia}}]{Mel:Mob:Zia:16}%
	\BibitemOpen
	\bibfield  {author} {\bibinfo {author} {\bibfnamefont {A.}~\bibnamefont
			{Mellor}}, \bibinfo {author} {\bibfnamefont {M.}~\bibnamefont {Mobilia}}, \
		and\ \bibinfo {author} {\bibfnamefont {R.}~\bibnamefont {Zia}},\ }\href
	{\doibase 10.1209/0295-5075/113/48001} {\bibfield  {journal} {\bibinfo
			{journal} {EPL (Europhysics Letters)}\ }\textbf {\bibinfo {volume} {113}},\
		\bibinfo {pages} {48001} (\bibinfo {year} {2016})}\BibitemShut {NoStop}%
	\bibitem [{\citenamefont {Mellor}\ \emph {et~al.}(2017)\citenamefont {Mellor},
		\citenamefont {Mobilia},\ and\ \citenamefont {Zia}}]{Mel:Mob:Zia:17}%
	\BibitemOpen
	\bibfield  {author} {\bibinfo {author} {\bibfnamefont {A.}~\bibnamefont
			{Mellor}}, \bibinfo {author} {\bibfnamefont {M.}~\bibnamefont {Mobilia}}, \
		and\ \bibinfo {author} {\bibfnamefont {R.}~\bibnamefont {Zia}},\ }\href
	{\doibase 10.1103/PhysRevE.95.012104} {\bibfield  {journal} {\bibinfo
			{journal} {Physical Review E}\ }\textbf {\bibinfo {volume} {95}},\ \bibinfo
		{pages} {012104} (\bibinfo {year} {2017})}\BibitemShut {NoStop}%
	\bibitem [{\citenamefont {J{\k{e}}drzejewski}(2017)}]{Jed:17}%
	\BibitemOpen
	\bibfield  {author} {\bibinfo {author} {\bibfnamefont {A.}~\bibnamefont
			{J{\k{e}}drzejewski}},\ }\href {\doibase 10.1103/PhysRevE.95.012307}
	{\bibfield  {journal} {\bibinfo  {journal} {Phys Rev. E}\ }\textbf {\bibinfo
			{volume} {95}},\ \bibinfo {pages} {012307} (\bibinfo {year}
		{2017})}\BibitemShut {NoStop}%
	\bibitem [{\citenamefont {Nyczka}\ and\ \citenamefont
		{Sznajd-Weron}(2013)}]{Nyc:Szn:13}%
	\BibitemOpen
	\bibfield  {author} {\bibinfo {author} {\bibfnamefont {P.}~\bibnamefont
			{Nyczka}}\ and\ \bibinfo {author} {\bibfnamefont {K.}~\bibnamefont
			{Sznajd-Weron}},\ }\href {\doibase 10.1007/s10955-013-0701-4} {\bibfield
		{journal} {\bibinfo  {journal} {J. Stat. Phys.}\ }\textbf {\bibinfo {volume}
			{151}},\ \bibinfo {pages} {174} (\bibinfo {year} {2013})}\BibitemShut
	{NoStop}%
	\bibitem [{\citenamefont {Vieira}\ and\ \citenamefont
		{Anteneodo}(2018)}]{Vie:Ant:18}%
	\BibitemOpen
	\bibfield  {author} {\bibinfo {author} {\bibfnamefont {A.~R.}\ \bibnamefont
			{Vieira}}\ and\ \bibinfo {author} {\bibfnamefont {C.}~\bibnamefont
			{Anteneodo}},\ }\href {\doibase 10.1103/PhysRevE.97.052106} {\bibfield
		{journal} {\bibinfo  {journal} {Physical Review E}\ }\textbf {\bibinfo
			{volume} {97}},\ \bibinfo {pages} {052106} (\bibinfo {year}
		{2018})}\BibitemShut {NoStop}%
	\bibitem [{\citenamefont {Nyczka}\ \emph {et~al.}(2018)\citenamefont {Nyczka},
		\citenamefont {Byrka}, \citenamefont {Nail},\ and\ \citenamefont
		{Sznajd-Weron}}]{Nyc:etal:18}%
	\BibitemOpen
	\bibfield  {author} {\bibinfo {author} {\bibfnamefont {P.}~\bibnamefont
			{Nyczka}}, \bibinfo {author} {\bibfnamefont {K.}~\bibnamefont {Byrka}},
		\bibinfo {author} {\bibfnamefont {P.~R.}\ \bibnamefont {Nail}}, \ and\
		\bibinfo {author} {\bibfnamefont {K.}~\bibnamefont {Sznajd-Weron}},\ }\href
	{\doibase 10.1371/journal.pone.0209620} {\bibfield  {journal} {\bibinfo
			{journal} {PLoS ONE}\ }\textbf {\bibinfo {volume} {13}} (\bibinfo {year}
		{2018}),\ 10.1371/journal.pone.0209620}\BibitemShut {NoStop}%
	\bibitem [{\citenamefont {Vieira}\ \emph {et~al.}(2020)\citenamefont {Vieira},
		\citenamefont {Peralta}, \citenamefont {Toral}, \citenamefont {San~Miguel},\
		and\ \citenamefont {Anteneodo}}]{Vie:etal:20}%
	\BibitemOpen
	\bibfield  {author} {\bibinfo {author} {\bibfnamefont {A.}~\bibnamefont
			{Vieira}}, \bibinfo {author} {\bibfnamefont {A.~F.}\ \bibnamefont {Peralta}},
		\bibinfo {author} {\bibfnamefont {R.}~\bibnamefont {Toral}}, \bibinfo
		{author} {\bibfnamefont {M.}~\bibnamefont {San~Miguel}}, \ and\ \bibinfo
		{author} {\bibfnamefont {C.}~\bibnamefont {Anteneodo}},\ }\href@noop {}
	{\bibfield  {journal} {\bibinfo  {journal} {arXiv:2002.04715v1}\ } (\bibinfo
		{year} {2020})}\BibitemShut {NoStop}%
	\bibitem [{\citenamefont {Nowak}\ and\ \citenamefont
		{Sznajd-Weron}(2019)}]{Now:Szn:19}%
	\BibitemOpen
	\bibfield  {author} {\bibinfo {author} {\bibfnamefont {B.}~\bibnamefont
			{Nowak}}\ and\ \bibinfo {author} {\bibfnamefont {K.}~\bibnamefont
			{Sznajd-Weron}},\ }\href {\doibase 10.1155/2019/5150825} {\bibfield
		{journal} {\bibinfo  {journal} {Complexity}\ }\textbf {\bibinfo {volume}
			{2019}} (\bibinfo {year} {2019}),\ 10.1155/2019/5150825}\BibitemShut
	{NoStop}%
	\bibitem [{\citenamefont {Nyczka}\ \emph
		{et~al.}(2012{\natexlab{b}})\citenamefont {Nyczka}, \citenamefont
		{Cis{\l}o},\ and\ \citenamefont {Sznajd-Weron}}]{Nyc:Cis:Szn:12}%
	\BibitemOpen
	\bibfield  {author} {\bibinfo {author} {\bibfnamefont {P.}~\bibnamefont
			{Nyczka}}, \bibinfo {author} {\bibfnamefont {J.}~\bibnamefont {Cis{\l}o}}, \
		and\ \bibinfo {author} {\bibfnamefont {K.}~\bibnamefont {Sznajd-Weron}},\
	}\href {\doibase 10.1016/j.physa.2011.07.050} {\bibfield  {journal} {\bibinfo
			{journal} {Physica A: Statistical Mechanics and its Applications}\ }\textbf
		{\bibinfo {volume} {391}},\ \bibinfo {pages} {317} (\bibinfo {year}
		{2012}{\natexlab{b}})}\BibitemShut {NoStop}%
	\bibitem [{\citenamefont {Peralta}\ \emph
		{et~al.}(2018{\natexlab{a}})\citenamefont {Peralta}, \citenamefont {Carro},
		\citenamefont {San~Miguel},\ and\ \citenamefont {R}}]{Per:etal:18}%
	\BibitemOpen
	\bibfield  {author} {\bibinfo {author} {\bibfnamefont {A.}~\bibnamefont
			{Peralta}}, \bibinfo {author} {\bibfnamefont {A.}~\bibnamefont {Carro}},
		\bibinfo {author} {\bibfnamefont {M.}~\bibnamefont {San~Miguel}}, \ and\
		\bibinfo {author} {\bibfnamefont {T.}~\bibnamefont {R}},\ }\href {\doibase
		10.1063/1.5030112} {\bibfield  {journal} {\bibinfo  {journal} {Chaos}\
		}\textbf {\bibinfo {volume} {28}},\ \bibinfo {pages} {075516} (\bibinfo
		{year} {2018}{\natexlab{a}})}\BibitemShut {NoStop}%
	\bibitem [{\citenamefont {Scheffer}\ \emph {et~al.}(2003)\citenamefont
		{Scheffer}, \citenamefont {Westley},\ and\ \citenamefont
		{Brock}}]{Sch:Wes:Bro:03}%
	\BibitemOpen
	\bibfield  {author} {\bibinfo {author} {\bibfnamefont {M.}~\bibnamefont
			{Scheffer}}, \bibinfo {author} {\bibfnamefont {F.}~\bibnamefont {Westley}}, \
		and\ \bibinfo {author} {\bibfnamefont {W.}~\bibnamefont {Brock}},\ }\href
	{\doibase 10.1007/s10021-002-0146-0} {\bibfield  {journal} {\bibinfo
			{journal} {Ecosystems}\ }\textbf {\bibinfo {volume} {6}},\ \bibinfo {pages}
		{493} (\bibinfo {year} {2003})}\BibitemShut {NoStop}%
	\bibitem [{\citenamefont {Vallacher}\ \emph {et~al.}(2017)\citenamefont
		{Vallacher}, \citenamefont {Nowak},\ and\ \citenamefont
		{Read}}]{Val:Now:Rea:17}%
	\BibitemOpen
	\bibfield  {author} {\bibinfo {author} {\bibfnamefont {R.}~\bibnamefont
			{Vallacher}}, \bibinfo {author} {\bibfnamefont {A.}~\bibnamefont {Nowak}}, \
		and\ \bibinfo {author} {\bibfnamefont {S.}~\bibnamefont {Read}},\ }\href
	{\doibase 10.4324/9781315173726} {\emph {\bibinfo {title} {Computational
				social psychology}}}\ (\bibinfo {year} {2017})\ pp.\ \bibinfo {pages}
	{1--381}\BibitemShut {NoStop}%
	\bibitem [{\citenamefont {Centola}\ \emph {et~al.}(2018)\citenamefont
		{Centola}, \citenamefont {Becker}, \citenamefont {Brackbill},\ and\
		\citenamefont {Baronchelli}}]{Cen:etal:18}%
	\BibitemOpen
	\bibfield  {author} {\bibinfo {author} {\bibfnamefont {D.}~\bibnamefont
			{Centola}}, \bibinfo {author} {\bibfnamefont {J.}~\bibnamefont {Becker}},
		\bibinfo {author} {\bibfnamefont {D.}~\bibnamefont {Brackbill}}, \ and\
		\bibinfo {author} {\bibfnamefont {A.}~\bibnamefont {Baronchelli}},\ }\href
	{\doibase 10.1126/science.aas8827} {\bibfield  {journal} {\bibinfo  {journal}
			{Science}\ }\textbf {\bibinfo {volume} {360}},\ \bibinfo {pages} {1116}
		(\bibinfo {year} {2018})}\BibitemShut {NoStop}%
	\bibitem [{\citenamefont {Watts}\ and\ \citenamefont
		{Strogatz}(1998)}]{Watt:Str:98}%
	\BibitemOpen
	\bibfield  {author} {\bibinfo {author} {\bibfnamefont {D.~J.}\ \bibnamefont
			{Watts}}\ and\ \bibinfo {author} {\bibfnamefont {S.~H.}\ \bibnamefont
			{Strogatz}},\ }\href {\doibase 10.1038/30918} {\bibfield  {journal} {\bibinfo
			{journal} {Nature}\ }\textbf {\bibinfo {volume} {393}},\ \bibinfo {pages}
		{440} (\bibinfo {year} {1998})}\BibitemShut {NoStop}%
	\bibitem [{\citenamefont {Abramiuk}\ and\ \citenamefont
		{Sznajd-Weron}(2020)}]{Abr:Szn:20}%
	\BibitemOpen
	\bibfield  {author} {\bibinfo {author} {\bibfnamefont {A.}~\bibnamefont
			{Abramiuk}}\ and\ \bibinfo {author} {\bibfnamefont {K.}~\bibnamefont
			{Sznajd-Weron}},\ }\href {\doibase 10.3390/e22010120} {\bibfield  {journal}
		{\bibinfo  {journal} {Entropy}\ }\textbf {\bibinfo {volume} {22}} (\bibinfo
		{year} {2020}),\ 10.3390/e22010120}\BibitemShut {NoStop}%
	\bibitem [{\citenamefont {Gleeson}(2013)}]{Gle:13}%
	\BibitemOpen
	\bibfield  {author} {\bibinfo {author} {\bibfnamefont {J.~P.}\ \bibnamefont
			{Gleeson}},\ }\href {\doibase 10.1103/PhysRevX.3.021004} {\bibfield
		{journal} {\bibinfo  {journal} {Phys. Rev. X}\ }\textbf {\bibinfo {volume}
			{3}},\ \bibinfo {pages} {021004} (\bibinfo {year} {2013})}\BibitemShut
	{NoStop}%
	\bibitem [{\citenamefont {Peralta}\ \emph
		{et~al.}(2018{\natexlab{b}})\citenamefont {Peralta}, \citenamefont {Carro},
		\citenamefont {San~Miguel},\ and\ \citenamefont {Toral}}]{Per:etal:18a}%
	\BibitemOpen
	\bibfield  {author} {\bibinfo {author} {\bibfnamefont {A.}~\bibnamefont
			{Peralta}}, \bibinfo {author} {\bibfnamefont {A.}~\bibnamefont {Carro}},
		\bibinfo {author} {\bibfnamefont {M.}~\bibnamefont {San~Miguel}}, \ and\
		\bibinfo {author} {\bibfnamefont {R.}~\bibnamefont {Toral}},\ }\href
	{\doibase 10.1088/1367-2630/aae7f5} {\bibfield  {journal} {\bibinfo
			{journal} {New Journal of Physics}\ }\textbf {\bibinfo {volume} {20}},\
		\bibinfo {pages} {103045} (\bibinfo {year} {2018}{\natexlab{b}})}\BibitemShut
	{NoStop}%
	\bibitem [{\citenamefont {Baronchelli}\ and\ \citenamefont
		{Pastor-Satorras}(2010)}]{Bar:Pas:10}%
	\BibitemOpen
	\bibfield  {author} {\bibinfo {author} {\bibfnamefont {A.}~\bibnamefont
			{Baronchelli}}\ and\ \bibinfo {author} {\bibfnamefont {R.}~\bibnamefont
			{Pastor-Satorras}},\ }\href {\doibase 10.1103/PhysRevE.82.011111} {\bibfield
		{journal} {\bibinfo  {journal} {Physical Review E}\ }\textbf {\bibinfo
			{volume} {82}},\ \bibinfo {pages} {011111} (\bibinfo {year}
		{2010})}\BibitemShut {NoStop}%
	\bibitem [{\citenamefont {Dunbar}\ \emph {et~al.}(2015)\citenamefont {Dunbar},
		\citenamefont {Arnaboldi}, \citenamefont {Conti},\ and\ \citenamefont
		{Passarella}}]{Dun:etal:15}%
	\BibitemOpen
	\bibfield  {author} {\bibinfo {author} {\bibfnamefont {R.}~\bibnamefont
			{Dunbar}}, \bibinfo {author} {\bibfnamefont {V.}~\bibnamefont {Arnaboldi}},
		\bibinfo {author} {\bibfnamefont {M.}~\bibnamefont {Conti}}, \ and\ \bibinfo
		{author} {\bibfnamefont {A.}~\bibnamefont {Passarella}},\ }\href {\doibase
		10.1016/j.socnet.2015.04.005} {\bibfield  {journal} {\bibinfo  {journal}
			{Soc. Netw.}\ }\textbf {\bibinfo {volume} {43}},\ \bibinfo {pages} {39 }
		(\bibinfo {year} {2015})}\BibitemShut {NoStop}%
	\bibitem [{\citenamefont {Doering}\ \emph {et~al.}(2018)\citenamefont
		{Doering}, \citenamefont {Scharf}, \citenamefont {Moeller},\ and\
		\citenamefont {Pruitt}}]{Doe:etal:18}%
	\BibitemOpen
	\bibfield  {author} {\bibinfo {author} {\bibfnamefont {G.}~\bibnamefont
			{Doering}}, \bibinfo {author} {\bibfnamefont {I.}~\bibnamefont {Scharf}},
		\bibinfo {author} {\bibfnamefont {H.}~\bibnamefont {Moeller}}, \ and\
		\bibinfo {author} {\bibfnamefont {J.}~\bibnamefont {Pruitt}},\ }\href
	{\doibase 10.1038/s41559-018-0592-5} {\bibfield  {journal} {\bibinfo
			{journal} {Nature Ecology and Evolution}\ }\textbf {\bibinfo {volume} {2}},\
		\bibinfo {pages} {1298} (\bibinfo {year} {2018})}\BibitemShut {NoStop}%
	\bibitem [{\citenamefont {Asch}(1955)}]{Asch:55}%
	\BibitemOpen
	\bibfield  {author} {\bibinfo {author} {\bibfnamefont {S.~E.}\ \bibnamefont
			{Asch}},\ }\href {\doibase 10.1038/scientificamerican1155-31} {\bibfield
		{journal} {\bibinfo  {journal} {Sci. Am.}\ }\textbf {\bibinfo {volume}
			{193}},\ \bibinfo {pages} {31} (\bibinfo {year} {1955})}\BibitemShut
	{NoStop}%
\end{thebibliography}

\end{document}